\documentclass[journal,10pt]{IEEEtran}
\usepackage{amsmath}
\usepackage{amssymb}
\usepackage{amsfonts}
\usepackage{graphicx}
\usepackage{epsfig}
\usepackage{subfigure}
\usepackage{psfrag}
\usepackage{color}
\usepackage{url}
\usepackage{cite}
\begin{document}

\title{ Secrecy Wireless Information and Power Transfer in Fading Wiretap Channel}
\author{Hong Xing, Liang Liu, and Rui Zhang
%\thanks{Copyright (c) 2014 IEEE. Personal use of this material is permitted. However, permission to use this material for any other purposes must be obtained from the IEEE by sending a request to pubs-permissions@ieee.org.}
\thanks{This paper has been presented in part at IEEE International Conference on Communications (ICC), Sydney, Australia, June 10-14, 2014.}\vspace{-.25in}
\thanks{H. Xing is with the Centre for Telecommunications Research, King's College London (e-mail:hong.xing@kcl.ac.uk). This work was done when she was a visiting student with the Department of Electrical and Computer Engineering, National University of Singapore.}
\thanks{L. Liu is with the Department of Electrical and Computer Engineering, National University of Singapore (e-mail:liu\_liang@nus.edu.sg).}
\thanks{R. Zhang is with the Department of Electrical and Computer Engineering, National University of Singapore (e-mail:elezhang@nus.edu.sg). He is also
with the Institute for Infocomm Research, A*STAR, Singapore.}}

\maketitle

\begin{abstract}
Simultaneous wireless information and power transfer (SWIPT) has recently drawn significant interests for its dual use of radio signals to provide wireless data and energy access at the same time. However,  a challenging secrecy communication issue arises as the messages sent to the information receivers (IRs) may be eavesdropped by the energy receivers (ERs), which are presumed to harvest energy only from the received signals. To tackle this problem, we propose in this paper an artificial noise (AN) aided transmission scheme to facilitate the secrecy information transmission to IRs and yet meet the energy harvesting requirement for ERs, under the assumption that the AN can be cancelled at IRs but not at ERs. Specifically, the proposed  scheme splits the transmit power into two parts, to send the confidential message to the IR and an AN to interfere with the ER, respectively. Under a simplified three-node wiretap channel setup,  the transmit power allocations and power splitting ratios over fading channels are jointly optimized to minimize the outage probability for delay-limited secrecy information transmission, or to maximize the average rate for no-delay-limited secrecy information transmission, subject to a combination of average and peak power constraints at the transmitter as well as an average energy harvesting constraint at the ER. Both the secrecy outage probability minimization and average rate maximization problems are shown to be non-convex, for each of which we propose the optimal solution based on the dual decomposition as well as suboptimal solution based on the alternating optimization. Furthermore, two benchmark schemes are introduced for comparison where the AN is not used at the transmitter and the AN is used but cannot be cancelled by the IR, respectively. Finally, the performances of proposed schemes are evaluated by simulations in terms of various trade-offs for wireless (secrecy) information versus energy transmissions.
\end{abstract}
% IEEEtran.cls defaults to using nonbold math in the Abstract.
% This preserves the distinction between vectors and scalars. However,
% if the conference you are submitting to favors bold math in the abstract,
% then you can use LaTeX's standard command \boldmath at the very start
% of the abstract to achieve this. Many IEEE journals/conferences frown on
% math in the abstract anyway.

% no keywords

\IEEEpeerreviewmaketitle
\setlength{\baselineskip}{1\baselineskip}
\newtheorem{fact}{Fact}
\newtheorem{assumption}{Assumption}
\newtheorem{theorem}{\underline{Theorem}}[section]
\newtheorem{lemma}{\underline{Lemma}}[section]
\newtheorem{corollary}{\underline{Corollary}}[section]
\newtheorem{proposition}{\underline{Proposition}}[section]
\newtheorem{algorithm}{\underline{Algorithm}}
\newtheorem{definition}{\underline{Definition}}[section]
\newtheorem{example}{\underline{Example}}[section]
\newtheorem{remark}{\underline{Remark}}[section]
\newcommand{\mv}[1]{\mbox{\boldmath{$ #1 $}}}
\newcommand{\smv}[1]{\small{\mbox{\boldmath{$ #1 $}}}}

\section{Introduction}\label{sec:introduction}

\PARstart{R}{ecently}, there has been an upsurge of interests in radio signals enabled  simultaneous wireless information and power transfer (SWIPT) (see e.g. \cite{Zhang2013MIMO,Zhou2013SWIPT,Liu2013,Xu2013multiuser} and the references therein). A typical SWIPT system consists of one access point (AP) that has constant power supply and broadcasts wireless signals carrying both information and energy to a set of distributed user terminals. Among these users, some operate as the information receivers (IRs) to decode the information from received signals, while the others operate as the energy receivers (ERs) to harvest energy. To overcome the significant power loss due to attenuation over distance and yet meet the energy harvesting requirement of practical applications, in SWIPT systems the ERs are generally deployed relatively closer to the AP than the IRs. However, this gives rise to a challenging physical (PHY)-layer security issue \cite{Wyner1975,Cheong1978}, as ERs may easily eavesdrop the information sent to IRs if they do not harvest energy as presumed.

In a SWIPT system with secrecy information transmission to the IRs, there are two conflicting goals in the transmission design: the power of the received signal at the ER is desired to be made large for efficient energy harvesting, but also needs to be kept sufficiently small to prevent information eavesdropping. To resolve this conflict, in this paper we propose to split the transmit signal into two parts, with one part carrying the secrecy information for the IR and the other part carrying an artificial noise (AN) to interfere with the ER to prevent from eavesdropping, while the total signal power received at the ER can still be kept high to satisfy its energy harvesting requirement. Note that in the conventional secrecy communication setup without the energy harvesting consideration, AN has been widely applied to improve the secrecy transmission rates \cite{Goel2008,X.Zhou2010,Liao2011transmit_beamforming,Li2013Spatially}, where a fraction of the transmit power was allocated to send randomly generated noise signals to reduce the amount of information decodable by the eavesdroppers. In \cite{Liu2014Secrecy}, AN was first applied in a multiple-input single-output (MISO) SWIPT system, where the joint information and energy beamforming design at the transmitter was investigated to maximize the secrecy rate of the IR subject to individual harvested energy constraints of ERs, or to maximize the weighted sum-power harvested by ERs subject to a given secrecy rate constraint at the IR. However, \cite{Liu2014Secrecy} considered the additive white Gaussian noise (AWGN) channels, while the optimal AN-aided secrecy transmission design for SWIPT systems over fading channels has not yet been addressed in the literature, which motivates this work. It is also worth pointing out that although channel fading is traditionally regarded as a detrimental factor to the wireless channel capacity, it can be exploited to reduce the secrecy communication outage probability \cite{bloch2008wireless,Barros2006,Liang2008,Khalil2009Opportunistic,Gungor2013Secrecy} or improve the wireless channel secrecy capacity \cite{bloch2008wireless,Liang2008,Gopala2008,Khisti2008fading}. For the secrecy outage probability minimization for wireless fading channels with stringent transmission delay constraint, \cite{Liang2008} has derived the optimal power allocations in the fading broadcast channel with confidential messages assuming the channel state information known at the transmitter (CSIT). While for maximizing the ergodic secrecy capacity (ESC) of fading channels with no-delay-limited transmission, the corresponding optimal power and rate allocation strategies have been studied in \cite{Gopala2008}. However, existing results for fading wiretap channels cannot be directly applied in our new SWIPT setup due to the additional energy harvesting requirement for the ER (which may also play a role of eavesdropper).

\begin{figure}
\begin{center}
 \scalebox{0.48}{\includegraphics*{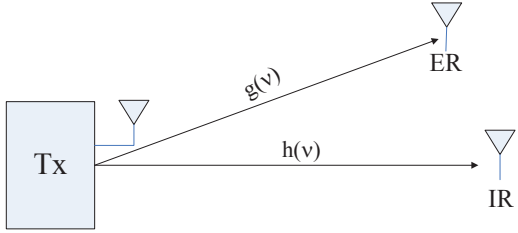}}
 \end{center}
\caption{The fading wiretap channel in a three-node SWIPT system.} \label{fig:fig1}
\end{figure}

In this paper, for the purpose of exposition we consider a three-node single-input single-output (SISO) fading wiretap channel consisting of one transmitter (Tx), one IR and one ER, each equipped with one antenna, as shown in Fig. \ref{fig:fig1}. We aim to minimize the outage probability for the IR for delay-limited secrecy transmission, or to maximize the ESC for the IR for no-delay-limited secrecy transmission, subject to the combined average and peak power constraints at the Tx as well as an average energy harvesting constraint at the ER. Note that unlike the existing literature on PHY-layer security, where the eavesdroppers are passive devices and thus their channels are practically assumed to  be unknown at the Tx, in this paper, we assume that the Tx knows the ER's eavesdropping channel since the ER needs to assist the Tx in obtaining its CSI to design the power allocations to satisfy its energy harvesting requirement. Moreover, for the AN-aided transmission, we assume that the Tx and IR both have the knowledge of the AN to be used prior to transmission via a known PHY-layer ``key'' distribution method \cite{Koorapaty1998,Koorapaty2000} (see Section \ref{sec:System Model} for the details); thus, the AN can be cancelled at the IR. However, the AN is kept strictly confidential to the ER and thus it cannot be cancelled at the ER. Such a scheme provides a theoretical upper-bound for the achievable secrecy rate of the SWIPT system under our consideration; whereas it is also worth noting that if the Tx and the IR are assumed to share certain common information {\it a priori}, our considered scheme may not be optimal as inspired by \cite{Khalil2009Opportunistic,Gungor2013Secrecy}. Nevertheless, we consider this scheme for its ease of implementation in practical SWIPT systems since the AN also plays the role of delivering wireless power to the ER (when it does not attempt to eavesdrop the information for the IR). Under this setup, we formulate first a secrecy outage probability minimization problem and then an ESC maximization problem,  for the three-node fading wiretap channel, which, however, are shown to be both non-convex. For each of the two problems, we first propose a dual decomposition based method to solve it optimally and then design an efficient suboptimal algorithm by iteratively optimizing the transmit power allocations and power splitting ratios over different fading states. For comparison, we also consider two benchmark schemes. In the first scheme, we assume that there is no AN employed at the Tx to facilitate the secrecy wireless information and power transfer, while in the second scheme, the AN is used but cannot be canceled by the IR. It is shown that the optimal power allocations for both schemes can be obtained based on the solution for the optimal scheme.

The remainder of the paper is organized as follows. Section \ref{sec:System Model} introduces the  SWIPT system model over a SISO fading wiretap channel. Section \ref{sec:Problem Formulation} presents the formulations of the proposed secrecy outage probability minimization problem and the ESC maximization problem. Section \ref{sec:Proposed Solutions to (P1)} and Section \ref{sec:Proposed Solutions to (P2)} propose both optimal and suboptimal solutions to the two formulated problems, respectively. Section \ref{sec:Benchmark Schemes} proposes two benchmark schemes and presents their optimal designs.  Section \ref{sec:Numerical Results} provides numerical results on the performance of various schemes proposed. Finally, Section \ref{sec:Conclusion} concludes the paper.

\section{System Model}\label{sec:System Model}
We consider the SISO fading wiretap channel for a three-node SWIPT system  as shown in Fig. \ref{fig:fig1}. It is assumed that there is one Tx, one IR and one ER, each equipped with one antenna. The complex channel coefficients from the Tx to IR and ER for
one particular fading state are denoted by \(u(\nu)\) and \(v(\nu)\), respectively, where \(\nu\) denotes the joint fading state. The power gains of the channels at fading state \(\nu\) are defined as \(h(\nu)=\vert u(\nu)\vert^2\) and \(g(\nu)=\vert v(\nu)\vert^2\); and it is assumed that at each fading state \(\nu\), both \(h(\nu)\) and \(g(\nu)\) are perfectly known at the Tx \footnote{\textcolor{black}{In practice, considering time division duplex (TDD) is used, at the beginning of each transmission block, the IR and ER can send their respective pilot signal to the Tx for it to estimate the reverse-link channel assuming short-term channel reciprocity between the Tx and IR/ER. TDD is also assumed for the subsequent description of secret ``key'' generation and transmission.}}
%\footnote{In practice, at the beginning of each transmission block, the Tx sends a pilot signal to the IR and ER and obtains the (downlink) CSI by collecting their respective channel estimation feedback.}.%For the ease of discussion in scenarios of wiretap channels, we define the channel between the Tx and the IR to be the main channel, and that between the Tx and the ER to be the eavesdropper (Eve)'s channel.
We further assume a block fading model such  that \(h(\nu) \) and \(g(\nu)\) remain constant during each block for each fading state \(\nu\), but can vary from block to block as \(\nu\) changes. It is assumed that \(h(\nu)\) and \(g(\nu)\) are two random variables with a continuous joint probability density function (pdf).

Since we are interested in secrecy information transmission to the IR, similar to \cite{Goel2008},  we assume that the transmit signal comprises of an information-bearing signal \(s_0\) and an AN-bearing signal \(s_1\). It is assumed that \(s_0\) is a circularly symmetric complex Gaussian (CSCG) random variable with zero mean and unit variance, denoted by  \(s_0\sim \mathcal{CN}(0,1)\). Furthermore, since \(s_1\) plays the role of AN to reduce the information eavesdropped by the ER and the worst case AN is known to be Gaussian distributed \cite{Goel2008}, we assume that \(s_1\) is also a CSCG random variable denoted by \(s_1\sim \mathcal{CN}(0,1)\), and is independent of $s_0$. The complex baseband transmit signal at fading state \(\nu\) is thus expressed as
\begin{align}
 x= \sqrt{(1-\alpha(\nu))p(\nu)}s_0+\sqrt{\alpha(\nu) p(\nu)}s_1\label{eq:transmit signal with AN},
\end{align}
where \(p(\nu)\) is the transmit power at fading state $\nu$ and \(0\leq\alpha(\nu)\leq 1\) denotes the portion of the transmit power allocated to the AN signal at fading state \(\nu\).  Moreover, similar to \cite{Liu2013}, in this paper we consider two types of power constraints on $p(\nu)$, namely, average power constraint (APC)
and peak power constraint (PPC). The APC
limits the average transmit power at the Tx over all fading states,
i.e., $E_{\nu}[p(\nu)]\leq P_{{\rm avg}}$, where $E_{\nu}[\cdot]$
denotes the expectation over $\nu$. In contrast, the PPC constrains
the instantaneous transmit power of the Tx at each fading state,
i.e., $p(\nu)\leq P_{{\rm peak}}$, $\forall \nu$. Without loss of generality, we assume $P_{{\rm avg}}\leq P_{{\rm peak}}$. The signals received at the IR and the ER are then respectively given by
\begin{align}
y_{{\rm IR}}&=u(\nu)x+n_{\rm IR} \nonumber \\ & =u(\nu)\left(\sqrt{(1-\alpha(\nu))p(\nu)}s_0+\sqrt{\alpha(\nu) p(\nu)}s_1\right)+n_{\rm IR}, \label{eq:received with AN 1}\\
y_{{\rm ER}} &=v(\nu)x+n_{\rm ER}  \nonumber \\ &  =v(\nu)\left(\sqrt{(1-\alpha(\nu))p(\nu)}s_0+\sqrt{\alpha(\nu) p(\nu)}s_1\right)+n_{\rm ER}, \label{eq:received with AN 2}
\end{align}where \(n_{\rm IR}\sim \mathcal{CN}(0,\sigma_1^2)\) and \(n_{\rm ER}\sim \mathcal{CN}(0,\sigma_2^2)\) denote the AWGN at the IR and the ER, respectively.

\textcolor{black}{As previously mentioned in the paper, it is assumed that the AN signal \(s_1\) is perfectly known to the IR (but not to the ER). A PHY-layer ``key'' distribution scheme with practical complexity is assumed for generating and cancelling the AN, which is described as follows. First, a large ensemble of seeds for a Gaussian pseudo-random generator are pre-stored at both the Tx and IR (but not available at the ER). We denote the index of each seed in the ensemble as a ``key'' in the sequel. Next, by randomly picking up one seed and transmitting its index to the IR before sending the confidential message at the beginning of each fading state, the Tx is able to generate a ``random'' AN sequence using the selected seed that is only known to the IR. Note that the ER does not have access to the seed ensemble; even if the ER attempts to decode the seed ensemble based on a long-term observation of the Tx-IR transmissions, the complexity is practically infeasible as the seed used at each fading state is random and unknown to the ER since the ``key'' (index of the seed in use) is also non-accessible by the ER. To achieve such secure ``key'' sharing, we further adopt a two-step phase-shift modulation based method \cite{Koorapaty1998,Koorapaty2000} by leveraging the short-term reciprocity of the wireless channels between the Tx and IR. Specifically, in the first step, the IR sends a pilot signal for the Tx to estimate the channel phase between the Tx and IR, while in the second step, the Tx randomly generates a seed index as a ``key'' and modulates it over the phase of the transmitted signal after pre-compensating the channel phase that it receives from the IR in the previous step. In this way, the IR is able to decode the ``key'' sent by the Tx from the received signal phases. Since the channel phase between the Tx and IR is different from that between the Tx/IR and ER, the ``key'' is secretly transmitted from the Tx to IR. Note that although the above ``key'' distribution method requires additional transmission time, it is negligible compared to the whole length of each transmission block if the channel coherence time is sufficiently large.}

With the above scheme, the associated interference at the IR in (\ref{eq:received with AN 1}), i.e., $u(\nu)\sqrt{\alpha(\nu) p(\nu)}s_1$, can be canceled at each fading state prior to decoding the desired information signal, \(s_0\). Then from (\ref{eq:received with AN 1}), the signal-to-noise ratio (SNR) at the IR at fading state \(\nu\) with a given pair of $\alpha(\nu)$ and $p(\nu)$ is expressed as
\begin{align}
 {\rm SNR}_{\rm IR}(\alpha(\nu),p(\nu))=\frac{(1-\alpha(\nu))h(\nu)p(\nu)}{\sigma_{\rm 1}^2}. \label{eq:SNR at the IR}
\end{align} Note that in practice the AN cancelation at the IR cannot be perfect, while the residue interference due to imperfect  AN cancellation could be included in the receiver noise power, i.e., \(\sigma_1^2\).
%Note that if the AN-bearing signal cannot be cancelled, the signal-to-interference-plus-noise ratio (SINR) at the IR at fading state \(\nu\)  is given by
%\begin{align}
%{\rm SINR}_{\rm IR}(\nu)=\frac{(1-\alpha(\nu))p(\nu)h(\nu)}{\alpha(\nu)p(\nu)h(\nu)+\sigma_R^2}.  \label{eq:SINR at the IR}
%\end{align}
 On the other hand, since the AN signal $s_1$ is assumed to be unknown to the ER and thus cannot be canceled, from (\ref{eq:received with AN 2}),  the SNR at the ER at fading state \(\nu\) is expressed as (assume that the ER eavesdrops  the information intended for the IR instead of harvesting energy)
\begin{align}
 {\rm SNR}_{\rm ER}(\alpha(\nu),p(\nu))=\frac{(1-\alpha(\nu))g(\nu)p(\nu)}{\alpha(\nu)g(\nu)p(\nu)+\sigma_2^2}. \label{eq:SINR at the ER}
\end{align}
Then, the achievable secrecy rate at fading state \(\nu\) can be expressed as \cite{Goel2008}
\begin{align}
R(\alpha(\nu),p(\nu))=&\bigg[\log_2\left(1+\frac{(1-\alpha(\nu))h(\nu)p(\nu)}{\sigma_1^2}\right) \nonumber \\ & -\log_2\left(1+\frac{(1-\alpha(\nu))g(\nu)p(\nu)}{\alpha(\nu)g(\nu)p(\nu)+\sigma_2^2}\right)\bigg]^+,\label{eq:mutual info AN-aided}
\end{align} where \([x]^+\triangleq\max(0,x)\).
%mutual info imperfect CSI
%R(\alpha(\nu),p(\nu))=&\bigg[\log_2\left(1+\frac{(1-\alpha(\nu))h(\nu)p(\nu)}{\alpha(\nu)p(\nu)\vert\delta u(\nu)\vert^2+\sigma_1^2}\right)\nonumber\\
%&-\log_2\left(1+\frac{(1-\alpha(\nu))g(\nu)p(\nu)}{\alpha(\nu)g(\nu)p(\nu)+\sigma_2^2}\right)\bigg]^+.
%where \( [x]^+\triangleq\max(0,x)\).

Next, for wireless power transfer, the amount of power harvested at fading state \(\nu\) at the ER is given by \cite{Zhang2013MIMO}
\begin{align}
  Q(p(\nu)) & =\zeta\left[(1-\alpha(\nu))g(\nu)p(\nu)+\alpha(\nu)g(\nu)p(\nu)\right] \nonumber\\
  & = \zeta g(\nu)p(\nu), \label{eq:harvested energy}
\end{align}
where \(0<\zeta\leq1\) denotes the energy harvesting efficiency. Note that the background noise power \(\sigma_2^2\) is ignored in \eqref{eq:harvested energy}, since it is typically very small as compared with the received signal power for energy harvesting. The average harvested power at the ER is thus  given by
\begin{align}
  Q_{\rm avg}=E_\nu\left[ Q(p(\nu))\right ]. \label{eq:ergodic harvested energy}
\end{align}

\section{Problem Formulation}\label{sec:Problem Formulation}
In this paper, we consider both delay-limited and no-delay-limited secrecy information transmission to the IR, for which the design problems are formulated in the following two subsections, respectively.

\subsection{Delay-Limited Secrecy Information Transmission}\label{subsec:Delay-Limited Secrecy Information Transmission}
First, consider the delay-limited secrecy information transmission to the IR, for which the outage probability is a relevant metric. Given a target rate \(r_0\), the secrecy outage probability at the IR can be expressed as \cite{Liang2008}
\begin{align}
  \delta=Pr(R(\alpha(\nu),p(\nu))<r_0), \label{eq:def of Pout}
\end{align}
where \(R(\alpha(\nu), p(\nu))\) is the achievable secrecy rate at fading state \(\nu\) given in \eqref{eq:mutual info AN-aided}, and \(Pr(\cdot)\) denotes the probability.  With CSIT, the transmitter-aware secrecy outage probability is generally minimized by the ``secrecy channel inversion'' based power allocation strategies \cite{Liang2008}. For convenience, we introduce the following indicator function for the event of outage with respect to the target secrecy rate \(r_0\) at each fading state \(\nu\):
 \begin{align}
  X(\nu)=\left\{\begin{array}{ll} 1\  & {\rm if}\, R(\alpha(\nu),p(\nu))<r_0, \\
  0 & {\rm otherwise.}\end{array}\right. \label{eq:indicator function}
\end{align}
It thus follows that the outage probability can be re-expressed as \(\delta=Pr(R(\alpha(\nu),p(\nu))<r_0)=E_\nu[X(\nu)]\).

For delay-limited secrecy information transmission, we aim at minimizing the secrecy outage probability for the IR by jointly optimizing the transmit power allocations, i.e., \(\{p(\nu)\}\), as well as the transmit power splitting ratios, i.e., \(\{\alpha(\nu)\}\) over different fading states, subject to a given pair of combined APC and PPC at the Tx, i.e., \(P_{\rm avg}\) and \(P_{\rm peak}\), as well as an average harvested power constraint at the ER, denoted by \(\bar Q\).  Therefore, we consider the following optimization problem.
\begin{align*}\mathrm{(P1)}:~\mathop{\mathtt{Minimize}}_{\{p(\nu), \alpha(\nu)\}}
& ~~~ E_\nu[X(\nu)]\\
\mathtt {Subject \ to}& ~~~E_\nu[p(\nu)]\leq P_{\rm avg}, \\
 & ~~~p(\nu)\leq P_{\rm peak}, \, \forall\nu,\\
 & ~~~  E_\nu[Q(p(\nu))]\geq \bar{Q},\\
& ~~~0\leq\alpha(\nu)\leq 1,\, \forall\nu.
\end{align*}

\subsection{No-Delay-Limited Secrecy Information Transmission}\label{subsec:No-Delay-Limited Secrecy Information Transmission}
Next, consider the no-delay-limited secrecy information transmission to the IR. In this case, ESC is a relevant metric that is expressed as
\begin{align}
{C_s}=E_\nu[R(\alpha(\nu),p(\nu))].\label{eq:def of ergodic capacity AN-aided}
\end{align}
With CSIT, \eqref{eq:def of ergodic capacity AN-aided} is generally maximized by the ``secrecy water-filling'' based power allocation policies \cite{Liang2008,Gopala2008}.

For no-delay-limited secrecy information transmission, we aim at maximizing the ESC for the IR subject to the same set of constraints (APC, PPC at the Tx, and an average harvested power constraint at the ER) as for the delay-limited case in (P1). Therefore, we consider the resulting optimization problem as follows.
 \begin{align*}\mathrm{(P2)}:~\mathop{\mathtt{Maximize}}_{\{ p(\nu), \alpha(\nu)\} }
& ~~~ E_\nu[R(\alpha(\nu),p(\nu))]\\
\mathtt {Subject \ to}& ~~~E_\nu[p(\nu)]\leq P_{\rm avg}, \\
 & ~~~p(\nu)\leq P_{\rm peak}, \, \forall\nu,\\
 & ~~~  E_\nu[Q(p(\nu))]\geq \bar{Q},\\
& ~~~0\leq\alpha(\nu)\leq 1,\, \forall\nu.
\end{align*}

Since the objective functions in (P1) and (P2) are in general non-convex and non-concave, respectively, (P1) and (P2) are non-convex problems. In the following two sections, we propose both optimal and suboptimal solutions to these two problems, respectively.
%We adopt a Rate-Energy (R-E) region \cite{Liu2013} metric, to compare the performances of varied schemes proposed towards the near optimal solution to (P1).  It is defined as the region comprising all the achievable secrecy rate \(R\) and the corresponding harvested energy \(E\), i.e., \(\{(R,E)\}\), for given \(P_{\rm avg}\) and \(P_{\rm peak}\). The R-E region of Problem (P1) is thus given as
%\begin{align}
%\label{eq:rate-energy region of (P1)}\mathcal{C}_{{\rm R-E}}\!\!\triangleq
%\bigcup_{\underset{p_1(\nu)\leq P_{\rm peak}}{E_\nu[p_1(\nu)]\leq P_{\rm avg}}} \!\!\bigg\{(R,E):R \leq C_s^\ast, E\leq \bar Q\bigg\},
%\end{align}
%where \(C_s^\ast\) is the optimal value of (P1) and \(\bar Q\) is the corresponding EH constraint given in (P1). Note that by firstly solving (P1) with specified \(Q=\bar{Q}\), \(C_s^\ast\) related to  \(\bar Q\) can be obtained. Next, by varying \(\bar Q\), the boundary of the R-E region, i.e., \(\left\{(R=C_s^\ast, E=\bar{Q})\right\}\) defined in \eqref{eq:rate-energy region of (P1)} can be thus characterized.

\section{Proposed Solutions to (P1) for Delay-Limited Case}\label{sec:Proposed Solutions to (P1)}
In this section, we propose both optimal and suboptimal solutions to (P1).
\subsection{Optimal Solution to (P1)}\label{subsec:Optimal Solution to (P1)}
First, we derive the optimal power allocations, i.e., \(\{p(\nu)\}\), and power splitting ratios, i.e., \(\{\alpha(\nu)\}\), to solve problem (P1). Following the similar analysis given in \cite{Liu2013}, under the assumption of continuous fading channel distributions,  (P1) can be shown to satisfy the ``time-sharing'' condition proposed in \cite{Yu2006}, and thus strong duality still approximately holds for this problem \cite{rockafellar1997convex}. Therefore,  we can apply the Lagrange duality method to solve (P1) optimally, as shown in the following.

The Lagrangian of (P1) is expressed as
\begin{align}
&L(\{p(\nu)\},\{\alpha(\nu)\},\lambda,\mu)\nonumber\\ &=E_\nu[X(\nu)]+\lambda(E_\nu[p(\nu)]-P_{\rm avg})-\mu(E_\nu[Q(p(\nu))]-\bar Q)\nonumber\\
& =E_\nu[X(\nu)+\lambda p(\nu)-\zeta\mu g(\nu)p(\nu)]-\lambda p_{\rm avg}+\mu\bar Q, \label{eq:Lagrangian function of (P1)}
\end{align}
 where \(\lambda\) and \(\mu\) are the dual variables associated with the APC, \(P_{\rm avg}\), and the average harvested power constraint, \(\bar Q\), respectively. Then the (partial) Lagrange dual function of (P1) is expressed as
 \begin{align}
   g(\lambda,\mu)=\!\!\!\min\limits_{\{p(\nu)\leq P_{\rm peak}\},\{\alpha(\nu)\in[0,1]\}}L(\{p(\nu)\},\{\alpha(\nu)\},\lambda,\mu). \label{eq:dual function of (P1)}
 \end{align}
 The dual problem of (P1) is thus given by
 \begin{align*}
\mathrm{(P1-dual)}:~\mathop{\mathtt{Maximize}}_{\lambda, \mu}
& ~~~ g(\lambda,\mu)\\
\mathtt {Subject \ to}& ~~~\lambda\geq 0,\;\mu\geq 0.
\end{align*}
 The minimization problem in \eqref{eq:dual function of (P1)} can be decoupled into parallel subproblems each for one fading state all having the same structure. Specifically, for one particular fading  state \(\nu\),  define \(L_1(p, \alpha)=X+\lambda p-\zeta\mu gp\). Then the associated subproblem given a pair of \(\lambda\) and \(\mu\) is expressed as
\begin{align*}
\mathrm{(P1-sub)}:~\mathop{\mathtt{Minimize}}_{p,\alpha}
& ~~~  L_1(p,\alpha)\\
\mathtt {Subject \ to}& ~~~p\leq P_{\rm peak},\\
& ~~~0\leq\alpha\leq 1.
\end{align*}
 Note that we have dropped the index \(\nu\) in \(p(\nu)\), \(\alpha(\nu)\) and \(X(\nu)\) for brevity.

Given any \(0\leq\alpha\leq1\), let \(p_1(\alpha)\) denote the minimum required power to maintain a target secrecy rate \(r_0\), i.e., \(R(\alpha,p)\ge r_0\),  it can be shown that
\begin{small}
\begin{align}
p_1(\alpha)=\left\{\begin{array}{ll} \frac{-(\alpha\sigma_1^2 g+(1-\alpha)\sigma_2^2h-2^{r_0}\sigma_1^2g)+\sqrt{\Delta}}{2\bar\alpha(1-\bar\alpha)hg}\ \!\!\!\!\!\!\!\!\!\!\!\!\!&{\rm if}\, 0<\alpha<1,\\
(2^{r_0}-1)\left. \middle/ \right. (\frac{h}{\sigma_1^2}-\frac{2^{r_0}g}{\sigma_2^2})\ {\rm if}\,\alpha=0\ \!\!\!\!\!&{\rm and}\ h>\tfrac{\sigma_1^2 2^{r_0}g}{\sigma_2^2},\\
+\infty \ \!\!\!\!\!\!\!\!\!\!\!\!\!&{\rm otherwise,}\end{array}\right. \label{eq:p2(bar alpha)}
\end{align}
\end{small}
where \(\Delta\) is given by
\begin{align}
\Delta &= \left(\alpha\sigma_1^2g+\sigma_2^2(1-\alpha)h \right)^2+2^{r_0}\big(2^{r_0}\sigma_1^4g^2-2\alpha\sigma_1^4g^2\nonumber\\
&=+(-4\alpha^2+6\alpha-2)\sigma_1^2\sigma_2^2hg \big).
 \label{eq:Delta for quadratic equation wrt p2}
\end{align}

Moreover, define \(\tilde\alpha\) as the optimal solution to the following problem:
\begin{align*}
\mathrm{(P1-search)}:~\mathop{\mathtt{Minimize}}_{\alpha}
& ~~~  p_1(\alpha)\\
\mathtt {Subject \ to}& ~~~0\leq\alpha\leq 1,
\end{align*}
which can be obtained by a simple one-dimension search. Then we have the following proposition.
\begin{proposition}
The optimal power allocations and power splitting ratios to problem (P1-sub) are given as
\begin{small}
\begin{align}
\!\!\!\!\!\!\!\left\{\!\!\!\begin{array}{ll} p^\ast=P_{\rm peak},\!\!\!\!&\alpha^\ast=\left\{\begin{array}{ll}\tilde\alpha &{\rm if}\,p_1(\tilde\alpha)\leq P_{\rm peak},\\0 & {\rm if}\,p_1(\tilde\alpha)>P_{\rm peak},\end{array}\right.  \ {\rm if}\  g>\frac{\lambda}{\zeta\mu} \\
p^\ast=p_1(\tilde\alpha),\!\!\!\!\!\!\!\!\!\!\!\!\!\!&\alpha^\ast=\tilde\alpha, \!\!\ {\rm if}\ g\leq\frac{\lambda}{\zeta\mu}\ {\rm and}\  p_1(\tilde\alpha)\leq\min \left(\frac{1}{\lambda-\zeta\mu g},\,P_{\rm peak} \right ), \\
p^\ast=0\, \!\!\!\!&\alpha^\ast=0, \ {\rm otherwise}.\end{array}\right. \label{eq:optimal solution for O-E region}
\end{align}\label{prop:optimal power control for O-E region}
\end{small}
\end{proposition}
\begin{IEEEproof}
  Please refer to Appendix \ref{proof of optimal power control for O-E region}.
\end{IEEEproof}

\begin{remark}
  We can draw some useful insight from Proposition \ref{prop:optimal power control for O-E region} for the optimal power control policy for a given pair of \((\lambda,\mu)\). When \(g>\frac{\lambda}{\zeta\mu}\), which means a relatively better channel condition for the ER, the Tx needs to transmit with peak power in order to maximize the harvested energy at the ER. Under this circumstance, if furthermore, \(p_1(\tilde\alpha)>P_{\rm peak}\), i.e.,  the outage event is inevitable, there is no need to optimize \(\alpha\) and thus it is set to be zero for simplicity; however, if \(p_1(\tilde\alpha)\leq P_{\rm peak}\), the outage can be avoided by setting \(\alpha\) to be any value satisfying \(p_1(\alpha)\leq P_{\rm peak}\), and thus we set \(\alpha=\tilde\alpha\). On the other hand, when \(g\leq \frac{\lambda}{\zeta\mu}\), we need to decide for the Tx whether to transmit with power \(p_1(\tilde\alpha)\) with power splitting ratio \(\tilde\alpha\), or to shut down its transmission to save power, based on whether \(p_1(\tilde\alpha)\) is smaller or larger than a certain threshold, i.e., \(\min(\frac{1}{\lambda-\zeta\mu g},P_{\rm peak})\).
\end{remark}

According to Proposition \ref{prop:optimal power control for O-E region}, with a given pair of \((\lambda,\mu)\), (P1-sub) can be efficiently solved state by state based on \eqref{eq:optimal solution for O-E region}. Problem (P1) is then iteratively solved by updating \((\lambda,\mu)\) via the ellipsoid method \cite{boyd2004convex}, for which the details are omitted for brevity. Notice that the required sub-gradient for updating \((\lambda,\mu)\) can be shown to be \(\left(E_\nu[p^\ast(\nu)]-P_{\rm avg}, \bar Q-E_\nu[Q(p^\ast(\nu))]\right)\), where \(p^\ast(\nu)\) is the optimal solution to problem (P1-sub) with given \(\lambda\) and \(\mu\).

\subsection{Suboptimal Solution to (P1)}\label{subsec:Suboptimal Solution to (P1)}
Note that the optimal solution given in Proposition \ref{prop:optimal power control for O-E region} requires an exhaustive search over \(\alpha\) in (P1-search) for \(\tilde\alpha\) in each of the fading states. In this subsection, we propose a suboptimal algorithm to solve (P1) with lower complexity based on the principle of alternating optimization. Specifically, by fixing \(\alpha(\nu)=\bar\alpha(\nu),\, \forall\nu\), we first optimize \(\{p(\nu)\}\) by solving the following problem.
\begin{align*}\mathrm{(P1.1)}:~\mathop{\mathtt{Minimize}}_{\{p(\nu)\}}
& ~~~E_\nu\left[X(\nu) \right ]\\
\mathtt {Subject \ to}& ~~~E_\nu[p(\nu)]\leq P_{\rm avg}, \\
& ~~~ p(\nu)\leq P_{\rm peak},\,\forall\nu,\\
 & ~~~ E_\nu[Q(p(\nu))]\geq \bar Q.
 \end{align*}

Let the optimal solution to (P1.1) be denoted by \(\{\bar p(\nu)\}\), with \(p(\nu)=\bar p(\nu),\, \forall\nu\), we then optimize \(\{\alpha(\nu)\}\) by solving the following problem.
\begin{align*}\mathrm{(P1.2)}:~\mathop{\mathtt{Minimize}}_{\{\alpha(\nu)\}}
& ~~~E_\nu\left[X(\nu) \right ]\\
\mathtt {Subject \ to}& ~~~0\leq \alpha(\nu)\leq 1,\, \forall\nu.
\end{align*}

The above procedure is repeated until both \(\{p(\nu)\}\) and \(\{\alpha(\nu)\}\) converge. In the following, we solve (P1.1) and (P1.2), respectively.

Problem (P1.1) is a non-convex problem since the objective function is not concave over \(p(\nu)\). However, similar to (P1), it satisfies the ``time-sharing'' condition, and thus we can use Lagrange duality method to solve it approximately with zero duality gap. Similarly as for problem (P1), problem (P1.1) can be decoupled into parallel subproblems each for one particular fading state and expressed as (by ignoring the fading state \(\nu\))
\begin{align*}
\mathrm{(P1.1-sub)}:~\mathop{\mathtt{Minimize}}_{p}
& ~~~  L_1(p)\\
\mathtt {Subject \ to}& ~~~p\leq P_{\rm peak},
\end{align*}
where \(L_1(p)=X+\lambda p-\zeta\mu gp\).

 Through the similar analysis as for Proposition \ref{prop:optimal power control for O-E region}, given any \(0\le\bar\alpha\le1\), the optimal solution to problem (P1.1-sub) is given as
 \begin{small}
\begin{align}
p^\ast= & \left\{\begin{array}{ll}P_{\rm peak}\  \!\!\!\!&{\rm if}\, g>\frac{\lambda}{\zeta\mu},\\
 p_1(\bar\alpha)\ \!\!\!\!&{\rm if}\, g\leq\frac{\lambda}{\zeta\mu}\ {\rm and}\ {p_1(\bar\alpha)\leq\min\left(\frac{1}{\lambda-\zeta\mu g},\,P_{\rm peak} \right )}, \\
0\ \!\!\!\!&{\rm otherwise}. \end{array}\right. \label{eq:optimal solution to (P1.1)}
\end{align}
\end{small}

With a given pair of \((\lambda,\mu)\), (P1.1-sub) can be efficiently solved state by state based on \eqref{eq:optimal solution to (P1.1)}. Problem (P1.1) can thus be iteratively solved by updating \((\lambda,\mu)\) via the ellipsoid method.

Next, we derive the optimal power splitting ratios \(\{\alpha(\nu)\}\) for problem (P1.2) with given \(\{\bar p(\nu)\}\). Note that the objective function of (P1.2) is separable over different fading states of \(\nu\). Hence, we only need to solve the following problem for each of the fading states.
\begin{equation}
\begin{split}\mathop{\mathtt{Minimize}}_{\alpha}
& ~~~ X\\
\mathtt {Subject \ to}& ~~~0\leq \alpha\leq 1.
\end{split} \label{eq:subproblem of (P1.2)}
\end{equation}
Note that we have dropped the index \(\nu\) for brevity.

Define \(\Phi=\{\alpha|R(\alpha,\bar p)\geq r_0\}\) as the set of \(\alpha\) that can guarantee the non-outage secrecy information transmission given \(\bar p\). If \(\Phi=\emptyset\), the outage cannot be avoided and thus any \(0\leq\alpha\leq 1\) can be the optimal solution to problem \eqref{eq:subproblem of (P1.2)}. Otherwise, any \(\alpha\in\Phi\) is optimal to problem \eqref{eq:subproblem of (P1.2)}. To select the best solution among the feasible \(\alpha\)'s, we solve the following problem.
\begin{align*}
\mathrm{(P1.2-sub)}:~\mathop{\mathtt{Maximize}}_{\alpha}
& ~~~ R(\alpha,\bar p)\\
\mathtt {Subject \ to}& ~~~0\leq \alpha\leq 1.
\end{align*}

Define \(x=\frac{\sigma_1^2}{h\bar p}-\frac{\sigma_2^2}{g\bar p}\). Then we have the following proposition.
\begin{proposition}
If \(\Phi\) is non-empty, the optimal solution  to problem (P1.2-sub) is given by
\begin{align}
\hat\alpha^\ast=\left\{\begin{array}{ll}0 \, & x< -1, \\
\frac{1}{2}+\frac{x}{2} \, & -1\leq x<1,\\
1\, & x\geq 1.
\end{array}\right.\label{eq:optimal solution to r(alpha)}
\end{align}\label{prop:optimal solution to (P1.2)}
\end{proposition}
\begin{IEEEproof}
  Please refer to Appendix \ref{proof of optimal power splitting ratio for O-E region}.
\end{IEEEproof}

By combining both the cases of $\Phi\neq\emptyset$ and $\Phi=\emptyset$, the optimal solution to problem (P1.2-sub) is given by \(\alpha^\ast=\hat\alpha^\ast\). Hence, problem (P1.2) for all \(\nu\)'s can be solved according to \eqref{eq:optimal solution to r(alpha)}.

With both problems (P1.1) and (P1.2) solved, we can then iteratively solve the two problems to obtain a suboptimal solution for (P1).  It is worth noting that the suboptimal algorithm proposed guarantees that the outage probability obtained is non-increasing after each iteration; thus the algorithm is ensured to at least converge to a locally optimal solution  to (P1).

%\begin{table}[htp]
%\begin{center}
%\caption{Suboptimal Algorithm I for (P1)} %\vspace{0.2cm}
% \hrule
%\vspace{0.2cm} \textbf{Algorithm }  \vspace{0.2cm}
%\hrule \vspace{0.2cm}
%\begin{itemize}
%\item {\bf Initialize} \(\{\alpha^{(0)}(\nu)\}\) and $i=0$
%\item {\bf Repeat} {\bf Set} $i=i+1$;
%\begin{itemize}
%\item [1)]  {\bf With} \(\{\bar{\alpha}(\nu)\}=\{\alpha^{(i-1)}(\nu)\}\) , {\bf obtain} \(\{p^{(i)}(\nu)\}\) by solving (P1.1) according to \eqref{eq:optimal solution to (P1.1)};
%\item [2)]  {\bf With} \(\{\bar{p}(\nu)\}=\{p^{(i)}(\nu)\}\), {\bf obtain} \(\{\alpha^{(i)}(\nu)\}\) by solving (P1.2) according to \eqref{eq:optimal solution to r(alpha)};
%\item [3)]{\bf Update} \(\delta^{(i)}=E_\nu[X^{(i)}(\nu)]\);
%\end{itemize}
%\item {\bf Until} \(\left\vert \delta^{(i)}-\delta^{(i-1)}\right\vert\leq\epsilon\), where  \(\epsilon\) is a small positive number that controls the algorithm accuracy.
%\end{itemize}
%\vspace{0.2cm} \hrule \label{algorithm:I} %\vspace{0.2 cm}
%\end{center}
%\end{table}

\section{Proposed Solutions to (P2) for No-Delay-Limited Case}\label{sec:Proposed Solutions to (P2)}
In this section, we  propose both optimal and suboptimal solutions to solve (P2).

\subsection{Optimal Solution to (P2)}\label{subsec:Optimal Algorithm II to (P2)}
First, we propose an optimal algorithm to solve (P2). Similar to Section \ref{subsec:Optimal Solution to (P1)}, based on the Lagrange duality method, problem (P2) can be decoupled into parallel subproblems all having the same structure and each for one fading state. Specifically, for one particular fading state $\nu$, we define
\(L_2(p,\alpha)=R(\alpha,p)-\lambda p +\zeta\mu gp\), where \(R(\alpha,p)\) is given in \eqref{eq:mutual info AN-aided}. Then the associated subproblem to solve for fading state $\nu$ is expressed as
\begin{align*}
\mathrm{(P2-sub)}:~\mathop{\mathtt{Maximize}}_{p,\alpha}
& ~~~  L_2(p,\alpha)\\
\mathtt {Subject \ to}& ~~~p\leq P_{{\rm peak}},\\
&~~~0\leq\alpha<1.
\end{align*}Note that we have dropped the index $\nu$ in $p(\nu)$ and $\alpha(\nu)$ for brevity.

 Since \(R(\alpha,p)\) is not concave over \(p\) and \(\alpha\), problem (P2-sub) is non-convex and thus difficult to be solved by applying convex optimization techniques. Hence, we propose a two-stage procedure to solve (P2-sub) optimally. First, we fix \(\alpha=\bar\alpha\) and then solve (P2-sub) to find the corresponding optimal power allocation \(\bar p\). Let \(f_\nu(\bar\alpha)\) denote the optimal value of  (P2-sub) given \(\alpha=\bar\alpha\). Next, the optimal \(\alpha^\ast\) to (P2-sub) is obtained by \(\max\limits_{0\leq\bar\alpha\leq1}\! f_\nu(\bar\alpha)\),  which can be solved by a one-dimension search over \(\bar\alpha\in[0,1]\). Therefore, in the following we focus on how to solve problem (P2-sub) with \(\alpha=\bar\alpha\).  First, we obtain the derivative of \(L_2(p,\bar\alpha)\) over \(p\) as
\begin{align}
\frac{\partial L_2(p,\bar\alpha)}{\partial p}  = \left\{\begin{array}{ll}\frac{Ap^3+Bp^2+Cp+D}{E} \, & {\rm if} \ p>\frac{\sigma_1^2}{\bar\alpha h}-\frac{\sigma_2^2}{\bar\alpha g}, \\
-\lambda +\mu\zeta g & {\rm otherwise},
\end{array}\right. \label{eq:derivative over p of (P2)}
\end{align}
where $A\triangleq \bar\alpha hg^2(\lambda-\mu\zeta g)(\bar\alpha-1)\ln2$, $B\triangleq h(\bar\alpha-1)F-\bar\alpha hg^2(\bar\alpha-1)-\bar\alpha g^2\sigma_1^2(\lambda-\mu\zeta g)\ln2$, $C\triangleq h\sigma_2^4(\lambda-\mu\zeta g)(\bar\alpha-1)\ln2-\sigma_1^2F-hg\sigma_2^2(\bar\alpha-1)^2-(hg\sigma_2^2+\bar\alpha hg\sigma_2^2)(\bar\alpha-1)$, $D\triangleq g\sigma_2^2\sigma_1^2(\bar\alpha-1)-h\sigma_2^4(\bar\alpha-1)-\sigma_2^4\sigma_1^2(\lambda-\mu\zeta g)\ln2$, $E \triangleq (\sigma_1^2+(1-\bar\alpha)ph)(\sigma_2^2+\bar\alpha pg)(\sigma_2^2+ pg)\ln 2$, and $F\triangleq g\sigma_2^2(\lambda-\mu\zeta g)(1+\bar\alpha)\ln2$. It can be observed from \eqref{eq:derivative over p of (P2)}  that the monotonicity of \(L_2(p,\bar\alpha)\) closely relates to the following cubic equation:
 \begin{align}
 Ap^3+Bp^2+Cp+D=0. \label{eq:cubic equation}
 \end{align}
 According to fundamental theorem of algebra, there are at most three roots (counted with multiplicity) to \eqref{eq:cubic equation}, denoted by \(x_1,x_2,\ {\rm and}, \ x_3\). Define a set as \(\mathcal{X}=\{x_1,x_2, x_3\}\). Since \(p\in\mathbb{R}\), only real roots in \(\mathcal{X}\) need to be taken into account. Thus, we define another set \(\Psi\) as follows:
\begin{align}
\Psi=\{x\vert x\in\mathbb{R}, 0\leq x\leq P_{\rm peak}, x\in\mathcal{X} \}\cup\{0,P_{\rm peak}\}, \label{eq:Psi}
\end{align}
 where \(2\leq\vert\Psi\vert\leq 5\), with \(\vert\cdot\vert\) denoting the cardinality of a set.  Note that \(\vert\Psi\vert=2\) when no real roots fall in the interval \([0,P_{\rm peak}]\),  while  \(\vert\Psi\vert=5\) when there are three distinct real roots  in \((0,P_{\rm peak})\). Next, it is easy to show that  the optimal \(p\) that maximizes  \(L_2(p, \bar\alpha) \) over \(p\in[0,P_{\rm peak}]\)  is obtained via a simple search over \(\Psi\), i.e.,
\begin{align}
\bar p(\lambda, \mu)=\arg\max\limits_{p\in\Psi} L_2(p, \bar\alpha). \label{eq:optimal solution to subproblem of (P2)}
\end{align}

 As a result, problem (P2-sub) is solved given any pair of \((\lambda,\mu)\). Problem (P2) is then solved by iteratively updating \((\lambda, \mu)\) by the ellipsoid method.

\subsection{Suboptimal Solution to (P2)}\label{subsec:Suboptimal Solution to (P2)}
Note that the optimal solution to (P2) requires a one-dimension search to find \(\alpha^\ast\) for each fading state.  Thus, in this subsection, we propose a suboptimal algorithm to solve (P2) with lower complexity based on alternating optimization. Specifically, by fixing \(\alpha(\nu)=\bar\alpha(\nu), \, \forall\nu\), we first optimize \(\{p(\nu)\}\) by solving the following problem.
\begin{align*}\mathrm{(P2.1)}:~\mathop{\mathtt{Maximize}}_{\{p(\nu)\}}
& ~~~E_\nu\left[R(\bar\alpha(\nu), p(\nu)) \right ]\\
\mathtt {Subject \ to}& ~~~E_\nu[p(\nu)]\leq P_{\rm avg}, \\
& ~~~ p(\nu)\leq P_{\rm peak},\,\forall\nu,\\
 & ~~~ E_\nu[Q(p(\nu))]\geq \bar Q.
 \end{align*}

Let the optimal solution of (P2.1) be denoted by \(\{\bar p(\nu)\}\). With \(p(\nu)=\bar p(\nu),\,\forall\nu\), we then optimize \(\{\alpha(\nu)\}\) by solving the following problem.
\begin{align*}\mathrm{(P2.2)}:~\mathop{\mathtt{Maximize}}_{\{\alpha(\nu)\}}
& ~~~E_\nu\left[R(\alpha(\nu),\bar p(\nu)) \right ]\\
\mathtt {Subject \ to}& ~~~0\leq \alpha(\nu)\leq 1,\, \forall\nu.
\end{align*}

The above two-stage procedure is repeated until both \(\{\bar p(\nu)\}\) and \(\{\bar\alpha(\nu)\}\) converge. In the following, we solve (P2.1) and (P2.2), respectively.

%\subsubsection{Optimal Solution to (P1.1)}\label{subsubsec:Optimal solution to (P1.1)}

 Similar to (P1.1), problem (P2.1) can be decoupled into parallel subproblems each for one fading state and expressed as (by ignoring the fading state \(\nu\))
\begin{align*}
\mathrm{(P2.1-sub)}:~\mathop{\mathtt{Maximize}}_{p}
& ~~~  L_2(p)\\
\mathtt {Subject \ to}& ~~~p\leq P_{\rm peak},
\end{align*}
where \(L_2(p)=R(\bar\alpha,p)-\lambda p+\zeta\mu gp\).

Note that problem (P2.1-sub) is equivalent to problem (P2-sub) with given \(\alpha=\bar{\alpha}\), the solution of which has been given in \eqref{eq:optimal solution to subproblem of (P2)}. As a result, problem (P2.1-sub) can be efficiently solved. Then, problem (P2.1) can be solved by iteratively updating \((\lambda, \mu)\) via  the ellipsoid method.

Next, we derive the optimal power splitting ratios \(\{\alpha(\nu)\}\) for problem (P2.2) with given $\{\bar{p}(\nu)\}$ obtained by solving  problem (P2.1).  Note that the objective function of (P2.2) is separable over different fading states. Thus, for each fading state \(\nu\), we need to solve the following problem (by dropping the index \(\nu\) for brevity):
\begin{align*}
\mathrm{(P2.2-sub)}:\mathop{\mathtt{Maximize}}_{\alpha}
& ~~~ R(\alpha,\bar p)\\
\mathtt {Subject \ to}& ~~~0\leq \alpha\leq 1.
\end{align*}
Note that problem (P2.2-sub) is the same as problem (P2.1-sub) in Section \ref{subsec:Suboptimal Solution to (P1)}, the solution of which has already been derived in Proposition \ref{prop:optimal solution to (P1.2)}. Hence, problem (P2.2) for all \(\nu\)'s can be solved according to \eqref{eq:optimal solution to r(alpha)}.

With both problems (P2.1) and (P2.2) solved, we can obtain a suboptimal solution for (P2) by iteratively solving these two problems. Similar to that for (P1), this suboptimal algorithm guarantees that the ESC is non-decreasing after each iteration, and thus convergence to at least a local optimal solution of (P2) is ensured.

%\begin{table}[htp]
%\begin{center}
%\caption{Suboptimal Algorithm II for (P2)} %\vspace{0.2cm}
%% \hrule
%%\vspace{0.2cm} \textbf{Algorithm }  \vspace{0.2cm}
%\hrule \vspace{0.2cm}
%\begin{itemize}
%\item {\bf Initialize} \(\{\alpha(\nu)^{(0)}\}\) and $i=0$
%\item {\bf Repeat} {\bf Set} $i=i+1$;
%\begin{itemize}
%\item [1)]  {\bf With} \(\{\bar{\alpha}(\nu)\}=\{\alpha^{(i-1)}(\nu)\}\) , {\bf obtain} \(\{p^{(i)}(\nu)\}\) by solving (P2.1) according to \eqref{eq:optimal solution to subproblem of (P2)};
%\item [2)]  {\bf With} \(\{\bar{p}(\nu)\}=\{p^{(i)}(\nu)\}\), {\bf obtain} \(\{\alpha^{(i)}(\nu)\}\) by solving (P2.2) according to \eqref{eq:optimal solution to r(alpha)};
%\item [3)]{\bf Update} \(C_s^{(i)}=E_\nu[R(\alpha^{(i)}(\nu),p^{(i)}(\nu))]\);
%\end{itemize}
%\item {\bf Until} \(\left\vert C_s^{(i)}-C_s^{(i-1)}\right\vert\leq\epsilon\), where  \(\epsilon\) is a small positive number that controls the algorithm accuracy.
%\end{itemize}
%\vspace{0.2cm} \hrule \label{algorithm:II} %\vspace{0.2 cm}
%\end{center}
%\end{table}

\section{Benchmark Schemes}\label{sec:Benchmark Schemes}
In this section, we introduce two benchmark schemes, where no AN is used at the transmitter, and the AN is used but is unknown to both the IR and ER, respectively.

First, consider the case when no AN is employed, i.e., \(\alpha(\nu)=0,\,\forall\nu\) for both the delay-limited secrecy transmission and the non-delay-limited counterpart. In this case, the SNRs at the IR and ER at fading state \(\nu\) given in (\ref{eq:SNR at the IR}) and (\ref{eq:SINR at the ER}) reduce to
\begin{align} & {\rm SNR}^\prime_{\rm IR}(\alpha(\nu),p(\nu))=\frac{h(\nu)p(\nu)}{\sigma_1^2}, \\ & {\rm SNR}^\prime_{\rm ER}(\alpha(\nu),p(\nu))=\frac{g(\nu)p(\nu)}{\sigma_2^2}, \end{align}respectively. Thus, the secrecy rate given in \eqref{eq:mutual info AN-aided} reduces to
\begin{align}
  &R^\prime(p(\nu))=\nonumber\\
  &\bigg[\log_2\left(1+\frac{h(\nu)p(\nu)}{\sigma_1^2}\right) -\log_2\left(1+\frac{g(\nu)p(\nu)}{\sigma_2^2}\right)\bigg]^+. \label{eq:mutual info NoAN}
\end{align}
It follows from \eqref{eq:mutual info NoAN} that the outage probability becomes \(\delta^\prime=Pr(R^\prime(p(\nu))<r_0)\), or equivalently, \(\delta^\prime=E_\nu[X^\prime(\nu)]\), where \(X^\prime(\nu)\) is modified from \eqref{eq:indicator function} as
 \begin{align}
  X^\prime(\nu)=\left\{\begin{array}{ll} 1\  & {\rm if}\, R^\prime(p(\nu))<r_0, \\
  0 & {\rm otherwise.}\end{array}\right. \label{eq:indicator function for (P1-NoAN)}
\end{align} Thus, (P1) reduces to the following problem.
 \begin{align*}\mathrm{(P1-NoAN)}:~\mathop{\mathtt{Minimize}}_{\{p(\nu)\}}
& ~~~ E_\nu[X^\prime(\nu)]\\
\mathtt {Subject \ to}& ~~~E_\nu[p(\nu)]\leq P_{\rm avg}, \\
 & ~~~p(\nu)\leq P_{\rm peak}, \, \forall\nu,\\
 & ~~~ E_\nu[Q(p(\nu))]\geq \bar{Q}.
\end{align*}

Accordingly, (P2) reduces to the following problem.
 \begin{align*}\mathrm{(P2-NoAN)}:~\mathop{\mathtt{Maximize}}_{\{p(\nu)\}}
& ~~~ E_\nu[  R^\prime(p(\nu))]\\
\mathtt {Subject \ to}& ~~~E_\nu[p(\nu)]\leq P_{\rm avg}, \\
& ~~~p(\nu)\leq P_{\rm peak},\, \forall\nu,\\
 & ~~~ E_\nu[Q(p(\nu))]\geq \bar{Q}.
\end{align*}

Note that (P1-NoAN) and (P2-NoAN) can be solved by simply setting \(\alpha(\nu)=0\) in (P1.1) and (P2.1), respectively.

Next, consider the case when the AN is used but is unknown to both the IR and ER, i.e., it cannot be canceled by the IR any more unlike that assumed in Sections \ref{sec:Proposed Solutions to (P1)} and \ref{sec:Proposed Solutions to (P2)}. In this case, the SNR expression at the ER at fading state $\nu$ is unchanged as (\ref{eq:SINR at the ER}), while the SNR at the IR at fading state $\nu$ needs to be modified as
\begin{align}{\rm SNR}^{\prime\prime}_{\rm IR}(\alpha(\nu),p(\nu))=\frac{(1-\alpha(\nu))h(\nu)p(\nu)}{\alpha(\nu)h(\nu)p(\nu)+\sigma_1^2}.\end{align}Then, the achievable secrecy rate given in \eqref{eq:mutual info AN-aided} is modified accordingly as
\begin{align}
R^{\prime\prime}(\alpha(\nu),p(\nu))=& \bigg[\log_2\left(1+\frac{(1-\alpha(\nu))h(\nu)p(\nu)}{\alpha(\nu)h(\nu)p(\nu)+\sigma_1^2}\right) \nonumber \\ & -\log_2\left(1+\frac{(1-\alpha(\nu))g(\nu)p(\nu)}{\alpha(\nu)g(\nu)p(\nu)+\sigma_2^2}\right)\bigg]^+.\label{eq:mutual info conventional}
\end{align}
 It follows from \eqref{eq:mutual info conventional} that the outage probability reduces to \(\delta^{\prime\prime}=Pr(R^{\prime\prime}(\alpha(\nu), p(\nu))<r_0)\), or equivalently, \(\delta^{\prime\prime}=E_\nu[X^{\prime\prime}(\nu)]\), where \(X^{\prime\prime}(\nu)\) is also modified from \eqref{eq:indicator function} as
 \begin{align}
  X^{\prime\prime}(\nu)=\left\{\begin{array}{ll} 1\  & {\rm if}\, R^{\prime\prime}(\alpha(\nu),p(\nu))<r_0, \\
  0 & {\rm otherwise.}\end{array}\right. \label{eq:indicator function for (P1-NoCancel)}
\end{align} Thus, (P1) is reformulated as
 \begin{align*}\mathrm{(P1-NoCancel)}:~\mathop{\mathtt{Minimize}}_{\{p(\nu), \alpha(\nu)\}}
& ~~~ E_\nu[X^{\prime\prime}(\nu)]\\
\mathtt {Subject \ to}& ~~~E_\nu[p(\nu)]\leq P_{\rm avg}, \\
 & ~~~p(\nu)\leq P_{\rm peak}, \, \forall\nu,\\
 & ~~~ E_\nu[Q(p(\nu))]\geq \bar{Q},\\
& ~~~ 0\leq\alpha(\nu)\leq 1,\, \forall\nu.
\end{align*}

 Accordingly, (P2) is reformulated as
\begin{align*}\mathrm{(P2-NoCancel)}:~\mathop{\mathtt{Maximize}}_{\{p(\nu),\alpha(\nu)\}}
& ~~~ E_\nu[ R^{\prime\prime}(\alpha(\nu),p(\nu))]\\
\mathtt {Subject \ to}& ~~~E_\nu[p(\nu)]\leq P_{\rm avg}, \\
& ~~~p(\nu)\leq P_{\rm peak},\, \forall\nu,\\
 & ~~~  E_\nu[Q(p(\nu))]\geq \bar{Q},\\
& ~~~0\leq\alpha(\nu)\leq 1,\, \forall\nu.
\end{align*}

(P1-NoCancel) and (P2-NoCancel) are both non-convex problems because \(X^{\prime\prime}(\nu)\) and \(R^{\prime\prime}(\alpha(\nu),p(\nu))\) are  non-convex and non-concave over \(p(\nu)\) and \(\alpha(\nu)\), respectively. However, we have the following proposition on their optimal solutions.
\begin{proposition}
The optimal solution to problem (P1-NoCancel) and (P2-NoCancel) must satisfy \(\alpha^\ast(\nu)=0,\forall\nu\). \label{prop:noCancel-optimal power splitting ratio}
\end{proposition}\begin{IEEEproof} Please refer to Appendix \ref{proof of noCancel-optimal power splitting ratio}. \end{IEEEproof}

Proposition \ref{prop:noCancel-optimal power splitting ratio} indicates that no AN should be used in (P1-NoCancel) or (P2-NoCancel), if it cannot be canceled by the IR. As a result, (P1-NoCancel) and (P2-NoCancel) are equivalent to the previous two problems, (P1-NoAN) and (P2-NoAN), respectively, which can be efficiently solved.

\section{Numerical Results}\label{sec:Numerical Results}
In this section, we provide numerical examples to evaluate the performance of our proposed optimal and suboptimal algorithms in Sections \ref{sec:Proposed Solutions to (P1)} and \ref{sec:Proposed Solutions to (P2)}, against the two benchmark schemes introduced in Section \ref{sec:Benchmark Schemes}. For comparison, we also consider the following heuristic approach to solve (P1) and (P2). First, we fix \(\alpha(\nu)=\bar\alpha,\,\forall\nu\), in (P1) or (P2), i.e., a uniform power splitting ratio for all fading states is assumed; then, we solve (P1.1) or (P2.1) to obtain the optimal \(\{p(\nu)\}\) . For convenience, in the sequel we refer to the above scheme as Fixed-\(\bar\alpha\). Compared with the two suboptimal algorithms proposed in Sections \ref{sec:Proposed Solutions to (P1)} and \ref{sec:Proposed Solutions to (P2)}, which require iteratively updating between \(\{\alpha(\nu)\}\)  and \(\{p(\nu)\}\) until their convergence, the algorithm of Fixed-\(\bar\alpha\) with fixed \(\alpha(\nu)=\bar\alpha, \forall\nu\), only needs one-shot for solving \(\{p(\nu)\}\), and thus has a much lower complexity.

We set \(P_{\rm avg}=100{\rm mW}\) or \(20\)dBm, \(P_{\rm peak}=1{\rm W}\)  or \(30{\rm dBm}\),  $\zeta=50\%$, and $\sigma_{1}^2=\sigma_{2}^2=-50$dBm. The distance-dependent pass loss model is given by
\begin{align}\label{eq:pass loss}
L=A_0\left(\frac{d}{d_0}\right)^{-\alpha}, d\geq d_0,
\end{align}where $A_0$ is set to be $10^{-3}$, $d$ denotes the distance between the Tx to the IR or ER, $d_0$ is a reference distance set to be $1$m, and $\alpha$ is the path loss exponent set to be $3$. It is assumed that $h(\nu)$ and $g(\nu)$ are independent exponentially distributed random variables (accounting for short-term Rayleigh fading) with their average power values specified by \eqref{eq:pass loss}.

\subsection{Secrecy Outage-Energy Trade-off}\label{subsec:Secrecy Outage-Energy Trade-off}
At first, we consider (P1) for characterizing the trade-offs between the secrecy outage probability for the IR and the average harvested power for the ER. Specifically, we adopt the (secrecy) Outage-Energy (O-E) region \cite{Liu2013}, which consists of all the pairs of achievable (secrecy) non-outage probability \(\epsilon\)  and average harvested power \(E\) for a given set of \(P_{\rm avg}\) and \(P_{\rm peak}\), which is defined as
\begin{align}
\mathcal{C}_{{\rm O-E}}\!\triangleq
\bigcup_{\underset{0\leq\alpha(\nu)\leq 1,\,\forall\nu}{\underset{p(\nu)\leq P_{\rm peak},\,\forall\nu}{E_\nu[p(\nu)]\leq P_{\rm avg}}}} \!\!\bigg\{(\epsilon,E):\epsilon \leq 1-\delta, E\leq  Q_{\rm avg}\bigg\},\label{eq:outage-energy region of (P1)}
\end{align}
where \(Q_{\rm avg}\) is given in \eqref{eq:ergodic harvested energy}, and \(1-\delta\) is the non-outage probability with respect to a given secrecy rate \(r_0\), where \(\delta\) is given in \eqref{eq:def of Pout}. Note that by solving (P1) with different \(\bar Q\)'s, the boundary of the corresponding O-E region for each considered scheme can be obtained accordingly.

\begin{figure}[ht]
\begin{center}
 \scalebox{0.58}{\includegraphics*{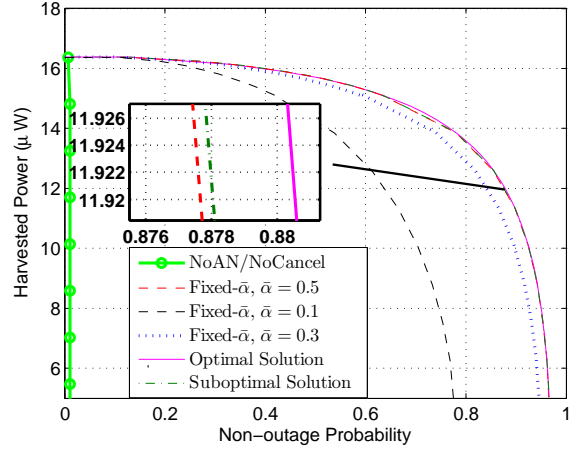}}
 \end{center}
\caption{Achievable O-E regions with a target secret rate \(r_0=6.5\)bits/sec/Hz by different power allocation schemes when the IR and ER are both $2$m away from the Tx.} \label{fig:fig2}
\vspace{-0.5ex}
\end{figure}

\begin{figure}[ht]
\begin{center}
 \scalebox{0.58}{\includegraphics*{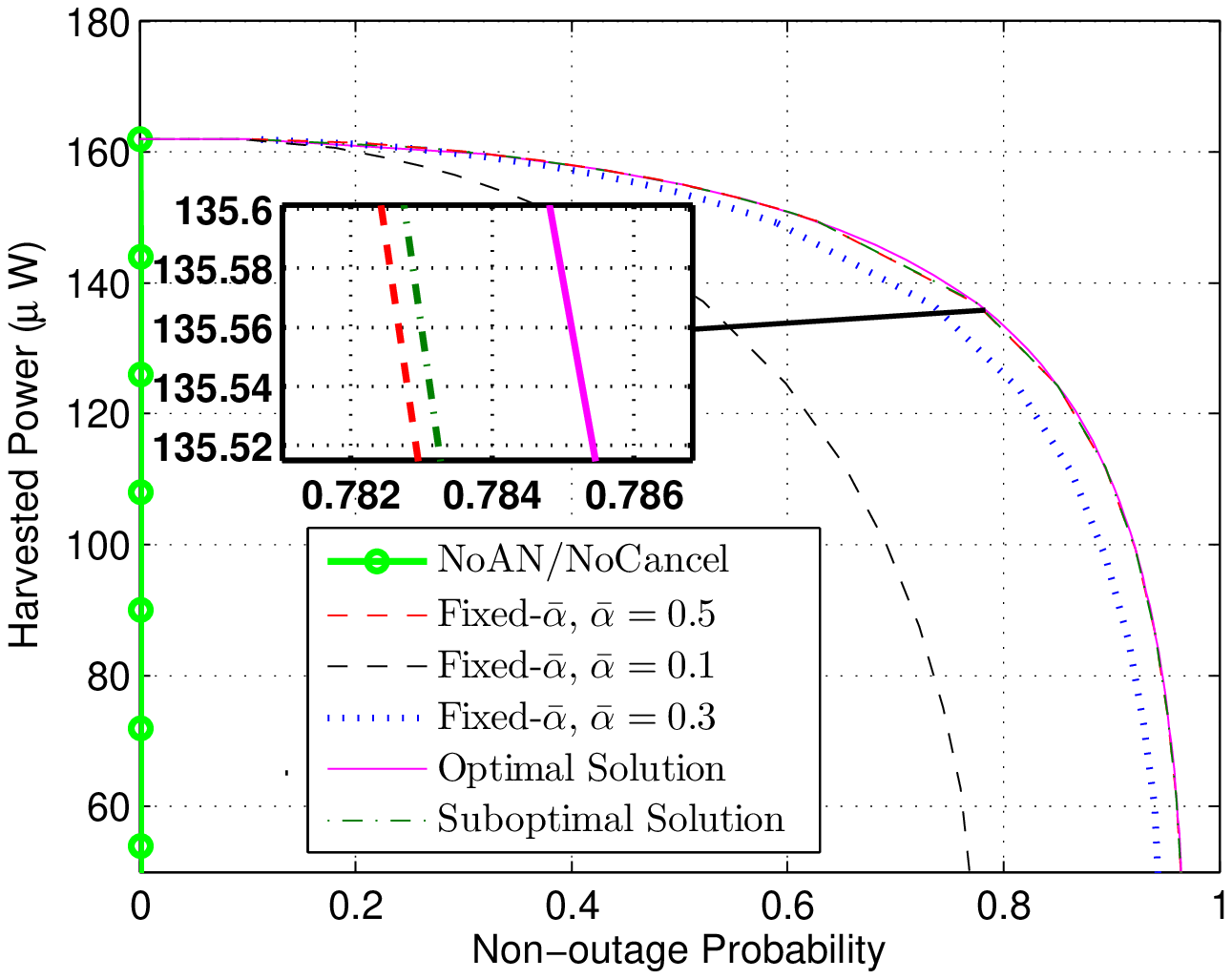}}
 \end{center}
\caption{Achievable O-E regions with a target secret rate \(r_0=6.5\)bits/sec/Hz by different power allocation schemes when the IR and ER are $2$m and $1$m away from the Tx, respectively.} \label{fig:fig3}
\vspace{-1.75ex}
\end{figure}

%\begin{figure}[ht]
%\begin{center}
% \scalebox{0.80}{\includegraphics*{Outage_varied_distance.eps}}
% \end{center}
%\caption{\textcolor{red}{Achieved non-outage probability with a target secrecy rate \(r_0=4\)bits/sec/Hz and average harvested power $\bar Q=2\mu$W versus the distance between the Tx and IR by different power allocation schemes, when the ER is fixed $2$m away from the Tx.}} \label{fig:fig3add}
%\end{figure}

 Consider a setup where the IR and the ER are of an identical distance of \(2\)m to the Tx. The target secret rate is set as \(r_0=6.5\)bps/Hz. Fig. \ref{fig:fig2} shows the O-E regions of the different schemes. It is observed that compared with both the schemes of NoAN and NoCancel, the proposed optimal algorithm with the use of AN achieves substantially improved O-E trade-offs thanks to the AN cancellation at the IR. For example, when an average harvested power of \(7.0\)\(\mu\)W is achieved, the secrecy outage probability can be made less than \(5\%\) versus more than \(98\%\). Furthermore, it is observed that when the AN can be canceled by the IR, the O-E region achieved by the suboptimal solution with alternating optimization is very close to that of the optimal solution. Furthermore, it is also observed that the O-E region achieved by Fixed-\(\bar\alpha\) with \(\bar\alpha=0.5,\forall\nu\),  has only negligible loss as compared to that of the optimal solution. The reason is as follows. In this setup, both the IR and the ER are very close to the Tx, and thus their average SNRs are high. It thus follows from \eqref{eq:optimal solution to r(alpha)} that when SNRs for the IR and the ER are high enough, \(x=\frac{\sigma_1^2}{h\bar p}-\frac{\sigma_2^2}{g\bar p}\) tends to be zero, and as a result, if the transmission is on, i.e., \(\bar p\neq0\),  the optimal power splitting ratios to (P1.2) becomes \(\alpha^\ast(\nu)\approx0.5,\,\forall\nu\). Last, we observe that the O-E trade-offs achieved by Fixed-\(\bar\alpha\) with other fixed values of \(\bar{\alpha}\) instead of  \(\bar\alpha=0.5\) deviate more notably from that of the optimal solution.

Next, we consider a more challenging setup for secrecy transmission when the ER is in more proximity to the Tx than the IR. Specifically, we assume that the IR and ER are \(2\)m and \(1\)m away from the Tx, respectively. Fig. \ref{fig:fig3} shows the O-E regions achieved by different schemes. Compared with Fig. \ref{fig:fig2}, it is observed that despite of the much worse channel condition for the IR than the ER, the achieved outage probability for secrecy transmission is almost unchanged. Also note from Fig. \ref{fig:fig3} that the achievable average harvested power for the ER is as about \(10\) times as that in Fig. \ref{fig:fig2}. However, it is observed that under this setup, the outage probability achieved by the schemes of NoAN or NoCancel is almost one due to the severely deteriorated average SNR of the IR's channel.

%\textcolor{red}{Finally, we evaluate how the achieved non-outage probability for secrecy transmission is compromised by the receiver-location based scheduling \cite{Zhang2013MIMO,Xu2013multiuser} of the IR and ER given a target secrecy rate \(r_0=4\)bits/sec/Hz for the IR and fixed average energy harvesting requirement \(\bar Q=2\mu\)W for the ER, where the ER and IR are located \(2\)m and from \(1\)m to \(10\)m distant away from the Tx, respectively. Except for the observation similar to Fig. \ref{fig:fig2} and Fig. \ref{fig:fig3} that the secrecy outage probability achieved by the suboptimal solution almost approaches that by the optimal solution and observably outperforms the heuristic scheme, it is also seen that the non-outage probability for secrecy transmission decreases with the distance that the IR is away from the Tx. Typically, when the IR is within \(4\)m away from the Tx, a secrecy outage of less than $10\%$ can be achieved by the optimal and suboptimal solution, whereas the delay-limited secrecy information transmission cannot be guaranteed with the IR's distance larger than $4$m away from the Tx. The reason can be intuitively explained as follows.  With the IR further away from the Tx and thus more deteriorated IR's channel, the given secrecy rate for each fading state is harder to be achieved, namely, the states with available power achieving secrecy transmission are less and less. }

\subsection{Secrecy Rate-Energy Trade-off}\label{subsec:Secrecy Rate-Energy Trade-off}

Next, we consider (P2) for characterizing the trade-offs between the ESC for the IR and the average harvested power for the ER. Specifically, we adopt the (secrecy) Rate-Energy (R-E) region \cite{Zhang2013MIMO}, which consists of all the pairs of achievable (secrecy) rate  \(R\) and harvested power \(E\) for a given set of \(P_{\rm avg}\) and \(P_{\rm peak}\), which is defined as
\begin{align}
\mathcal{C}_{{\rm R-E}}\!\triangleq
\bigcup_{\underset{0\leq\alpha(\nu)\leq 1}{\underset{p(\nu)\leq P_{\rm peak}}{E_\nu[p(\nu)]\leq P_{\rm avg}}}} \!\!\bigg\{(R,E):R \leq C_s, E\leq  Q_{\rm avg}\bigg\},\label{eq:rate-energy region of (P1)}
\end{align}
where \(Q_{\rm avg}\) is given in \eqref{eq:ergodic harvested energy}, and $C_s$ is expressed as $C_s=E_\nu[R(\nu)]$, with $R(\nu)$ given in (\ref{eq:mutual info AN-aided}), (\ref{eq:mutual info NoAN}) and (\ref{eq:mutual info conventional}), respectively, for different schemes. Note that by solving (P2) with different \(\bar{Q}\)'s, the boundary of the corresponding  R-E region for each considered scheme can be obtained.

\begin{figure}
\begin{center}
 \scalebox{0.58}{\includegraphics*{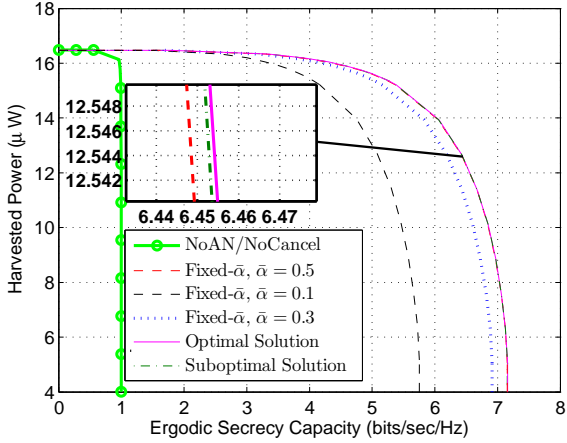}}
 \end{center}
\caption{Achievable R-E regions by different power allocation schemes when the IR and ER are both $2$m away from the Tx.} \label{fig:fig4}
\vspace{-1.25ex}
\end{figure}

\begin{figure}
\begin{center}
 \scalebox{0.58}{\includegraphics*{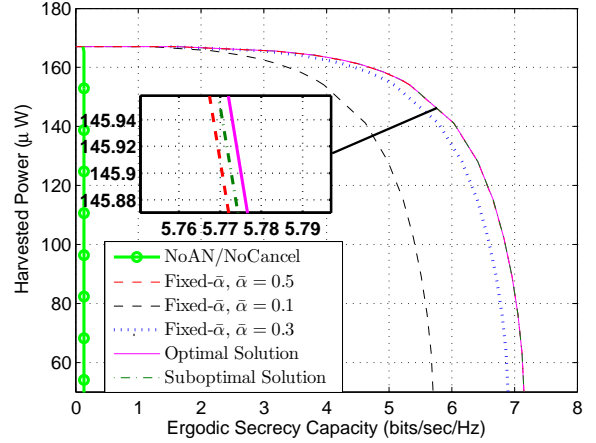}}
 \end{center}
\caption{Achievable R-E regions by different power allocation schemes when the IR and ER are $2$m and $1$m away from the Tx, respectively.} \label{fig:fig5}
\vspace{-2.00ex}
\end{figure}

%\begin{figure}[ht]
%\begin{center}
% \scalebox{0.80}{\includegraphics*{Rate_varied_distance.eps}}
% \end{center}
%\caption{\textcolor{red}{Achieved ergodic secrecy capacity with a target average harvested power $\bar Q=100\mu$W versus the distance between the Tx and IR by different power allocation schemes, when the ER is fixed $1$m away from the Tx.}} \label{fig:fig5add}
%\end{figure}

 Similar to the case of O-E region, we first consider the setup when the IR and the ER are of an identical distance of $2$m to the Tx. Fig. \ref{fig:fig4} shows the R-E regions of the different schemes. It is observed that compared with the scheme of NoAN (or NoCancel), the proposed AN-aided optimal solution achieves substantially improved R-E trade-offs due to the cancelable AN at the IR. For example, when an average harvested power of $6\mu$W is achieved, the ESC is increased by about $700\%$. Furthermore, it is observed that when the AN can be canceled by the IR, the R-E region achieved by the suboptimal solution is very close to that by the optimal solution. Finally, similar to the case of O-E region, the R-E region achieved by Fixed-\(\bar\alpha\) with \(\bar\alpha=0.5,\forall\nu\), is the best compared with those achieved by other fixed values of \( \bar{\alpha}\), i.e., \(\bar\alpha=0.1\) and \(\bar\alpha=0.3\).

Next, we consider the same setup with unequal distances from the Tx to the ER and IR as for Fig. \ref{fig:fig3}. Fig. \ref{fig:fig5} shows the R-E regions achieved by different schemes. Compared to Fig. \ref{fig:fig4}, it is observed that the performance gaps between the proposed optimal/suboptimal solutions and the scheme of NoAN or NoCancel become more substantial.

\section{Conclusion}\label{sec:Conclusion}
This paper studies the important issue of physical (PHY)-layer security in emerging simultaneous wireless information and power transfer (SWIPT) applications. Under a simplified three-node fading wiretap channel setup, we propose a dual use of the artificial noise (AN) for both interfering with and transferring energy to the ER, under the assumption that the AN is perfectly canceled at the IR. We jointly optimize the transmit  power allocations and power splitting ratios over the fading channel to minimize the outage probability for delay-limited secrecy transmission, and to maximize the average rate for no-delay-limited secrecy transmission, respectively, subject to the combined average and peak power constraint at the Tx, as well as an average energy harvesting constraint at the ER. We derive optimal solutions to these non-convex problems, and also propose suboptimal solutions of lower complexity based on the alternating optimization technique. Through extensive simulation results, we show that the proposed schemes achieve considerable (secrecy) Outage-Energy (O-E) and (secrecy) Rate-Energy (R-E) trade-off gains, as compared to the schemes without the use of AN.

% conference papers do not normally have an appendix

\appendices

\section{Proof of Proposition \ref{prop:optimal power control for O-E region}}\label{proof of optimal power control for O-E region}
We prove Proposition \ref{prop:optimal power control for O-E region} for the two cases of \(p_1(\tilde\alpha)>P_{\rm peak}\) and \(p_1(\tilde\alpha)\leq P_{\rm peak}\), respectively, shown as follows.
    \begin{itemize}
      \item[1)] {\bf Case I:} \(p_1(\tilde\alpha)>P_{\rm peak}\)

     In this case, since the minimum power for achieving \(r_0\) already exceeds \(P_{\rm peak}\), the outage is inevitable. Hence,
     \begin{align}
       L_1(p,\alpha)=1+(\lambda-\zeta\mu g)p. \label{eq:case I:Lagrangian}
     \end{align}
    To minimize \(L_1(p, \alpha)\), we have \begin{align}
p^\ast= & \left\{\begin{array}{ll} P_{\rm peak}\ &{\rm if}\, \lambda-\zeta\mu g<0 \\
 0\ &{\rm otherwise. }\end{array}\right. \label{eq:case I:optimal p}
\end{align} Note that since in this case \(X\equiv1\), \(\alpha\) can take any value over the interval \([0,1]\) and thus we set \(\alpha^\ast=0\) for convenience.

\item[2)] {\bf Case II:} \(p_1(\tilde\alpha)\leq P_{\rm peak}\)

In this case, the outage can be avoided by jointly optimizing \(p\) and \(\alpha\). As a result, we have
\begin{align}
L_1(p,\alpha)= & \left\{\begin{array}{ll} 1+(\lambda-\zeta\mu g)p\ &{\rm if}\, 0\leq p<p_1(\tilde\alpha), \\
 (\lambda-\zeta\mu g)p\ &{\rm if}\, p_1(\tilde\alpha)\leq p\leq P_{\rm peak.}\end{array}\right. \label{eq:case II:L}
\end{align}
According to \eqref{eq:case II:L}, the optimal power allocation to minimize \(L_1(p,\alpha)\) also depends on whether \(\lambda-\zeta\mu g< 0\) or not. Thus, in the following we further discuss two subcases.
\begin{itemize}
  \item  {\bf Subcase II-1:} \(\lambda-\zeta\mu g< 0\).
In this subcase, given any \(\alpha=\bar\alpha\) with \(p_1(\bar\alpha)\leq P_{\rm peak}\), \(L_1(p,\bar\alpha)\) is a monotonically decreasing function over \(p\). As a result, over the interval \(0\leq p\leq p_1(\bar\alpha)\), \(L_1(p,\bar\alpha)\) is minimized by \(p=p_1(\bar\alpha)\); while over the interval \(p_1(\bar\alpha)<p\leq P_{\rm peak}\), it is minimized by \(p=P_{\rm peak}\). Note that given any \(\bar\alpha\) with \(p_1(\bar\alpha)\leq P_{\rm peak}\), it follows that \(1+(\lambda-\zeta\mu g)p_1(\bar\alpha)>(\lambda-\zeta\mu g)P_{\rm peak}\). Therefore, the optimal power allocation for any \(\bar\alpha\) is \(p^\ast=P_{\rm peak}\). Moreover, any \(\bar\alpha\) that satisfies \(p_1(\bar\alpha)\leq P_{\rm peak}\) is optimal.

\item {\bf Subcase II-2:} \(\lambda-\zeta\mu g\geq 0\).
In this subcase, given any \(\alpha=\bar\alpha\) with \(p_1(\bar\alpha)\leq P_{\rm peak}\), \(L_1(p,\bar\alpha)\) is a monotonically increasing function over \(p\). As a result, over the interval \(0\leq p< p_1(\bar\alpha)\), \(L_1(p,\bar\alpha)\) is minimized by \(p=0\) (i.e., \(L_1^\ast(p,\bar\alpha)=1\)); while over the interval \(p_1(\bar\alpha)\leq p\leq P_{\rm peak}\), it is minimized by \(p=p_1(\bar\alpha)\). Furthermore, \(p_1(\bar\alpha)\) can be minimized by setting \(\bar\alpha=\tilde\alpha\) (i.e., \(L_1^\ast(p,\bar\alpha)=(\lambda-\zeta\mu g)p_1(\tilde\alpha)\)). Hence, the optimal power allocation for minimizing \(L_1(p,\bar\alpha)\) depends on the relationship between \(1\) and \((\lambda-\zeta\mu g)p_1(\tilde\alpha)\). If \(1<(\lambda-\zeta\mu g)p_1(\tilde\alpha)\), since \(p^\ast=0\), any \(\bar\alpha\) is optimal and thus we set \(\alpha^\ast=0\) for simplicity; however, if \(1\geq (\lambda-\zeta\mu g)p_1(\tilde\alpha)\), the optimal power allocation is \(p^\ast=p_1(\tilde\alpha)\) with the optimal power splitting ratio \(\alpha^\ast=\tilde\alpha\).
\end{itemize}
\end{itemize}

 By combing the above two cases of \(p_1(\tilde\alpha)>P_{\rm peak}\) and \(p_1(\tilde\alpha)\leq P_{\rm peak}\), Proposition \ref{prop:optimal power control for O-E region} is thus proved.

 %\subsection{Proof of Sub-gradient for Solving (P1) }\label{proof of sub-gradient for (P1)}
 %For (P1) with a dual objective given in \eqref{eq:dual function of (P1)}, given any \((\lambda^\prime, \mu^\prime)^T\succeq 0\), we have
 %\begin{align}
  % g(\lambda^\prime,\mu^\prime)\leq & L(\{p^\ast(\nu)\},\{\alpha^\ast(\nu)\},\lambda^\prime, \mu^\prime)\nonumber\\
%= & L(\{p^\ast(\nu)\},\{\alpha^\ast(\nu)\},\lambda, \mu)+(\lambda^\prime-\lambda)(E_\nu[p^\ast(\nu)]-P_{\rm avg})+(\mu^\prime-\mu)(\bar Q-E_\nu[Q(p^\ast(\nu))])\nonumber\\
%=&L(\{p^\ast(\nu)\},\{\alpha^\ast(\nu)\},\lambda, \mu)+(E_\nu[p^\ast(\nu)]-P_{\rm avg}, \bar Q-E_\nu[Q(p^\ast(\nu))])(\lambda^\prime-\lambda, \mu^\prime-\mu)^T. \label{eq:sub-gradient for g(lambda,mu)}
% \end{align}
% Thus, the sub-gradient for maximizing \eqref{eq:dual function of (P1)} in the ellipsoid method is proved to be \((E_\nu[p^\ast(\nu)]-P_{\rm avg}, \bar Q-E_\nu[Q(p^\ast(\nu))])\).

\section{Proof of Proposition \ref{prop:optimal solution to (P1.2)}} \label{proof of optimal power splitting ratio for O-E region}
According to \eqref{eq:mutual info AN-aided}, the derivative of  \(R(\alpha,\bar p)\) over \(\alpha\) is given by
\begin{align}
\frac{\partial R(\alpha, \bar p)}{\partial\alpha}=\left\{\begin{array}{ll}\frac{(1-2\alpha+x)hg\bar p^2}{\ln2(\sigma_1^2+(1-\alpha)h\bar p)(\sigma_2^2+\alpha g\bar p)} \, & {\rm if} \ \alpha \ge x,\\
0 \, & {\rm otherwise, }
\end{array}\right.  \label{eq:derivative over alpha of (P1.2)}
\end{align}
where \(x=\frac{\sigma_1^2}{h\bar{p}}-\frac{\sigma_2^2}{g\bar{p}}\). It can be shown from (\ref{eq:derivative over alpha of (P1.2)}) that if $x<-1$, then $\frac{\partial R(\alpha, \bar p)}{\partial\alpha}<0$ with $0\leq \alpha\leq 1$. Thus, $R(\alpha, \bar p)$ is a monotonically decreasing function over \(\alpha\) in the interval $[0,1]$, and the optimal solution to problem (\ref{eq:subproblem of (P1.2)}) is $\alpha^\ast=0$. If $-1\leq x<1$, it can be shown that $R(\alpha, \bar p)$ is a non-decreasing function of \(\alpha\) over the interval $[0,\frac{1}{2}+\frac{x}{2}]$, but a monotonically decreasing function over $(\frac{1}{2}+\frac{x}{2},1]$. As a result, we have $\alpha^\ast=\frac{1}{2}+\frac{x}{2}$. Finally, if $x\geq 1$, $\frac{\partial R(\alpha, \bar p)}{\partial\alpha}\geq 0$, and thus $R(\alpha, \bar p) $ is non-decreasing over $\alpha\in[0,1]$.  In this case, the optimal solution to problem (P1.2-sub) is $\alpha^\ast=1$. Proposition \ref{prop:optimal solution to (P1.2)} is thus proved.

%  \subsection{Proof of Corollary \ref{cor:noAN-optimal power control for O-E region}}\label{proof of noAN-optimal power control for O-E region}
%  The similar method for solving \eqref{eq:subproblem of (P1)} can be applied by substituting \(\alpha=0\). However, since the minimum power for maintaining the secrecy rate of \(r_0\) now becomes \(p_2^\prime=p_2(\bar\alpha)|_{\bar\alpha=0}=
%\frac{\sigma_1^2\sigma_2^2(2^{r_0}-1)}{\sigma_2^2h-\sigma_1^22^{r_0}g}\), we must have its denominator \(\sigma_2^2h-\sigma_1^22^{r_0}g>0\), i.e., \(g<\frac{\sigma_2^2h}{\sigma_1^22^{r_0}}\), to guarantee a positive \(p_2^\prime\). Therefore, the optimal power allocations in \eqref{eq:optimal solution for O-E region} must be modified into \eqref{eq:optimal power control for O-E region without AN} by taking \(g<\frac{\sigma_2^2h}{\sigma_1^22^{r_0}}\) into consideration.

\section{Proof of Proposition \ref{prop:noCancel-optimal power splitting ratio}}\label{proof of noCancel-optimal power splitting ratio}
 For problems (P1-NoCancel) and (P2-NoCancel), suppose that the average harvested power constraint is not present, the optimal power splitting ratios for both problems can be shown to be \(\alpha^\ast(\nu)=0,\forall\nu\), by solving \(\max\limits_{0\leq\alpha(\nu)\leq 1}\!R^{\prime\prime}(\alpha(\nu),\bar p(\nu))\) at each fading state \(\nu\) (c.f. \eqref{eq:mutual info conventional}), according to \cite{Liu2010MIMOGaussian_broadcast}.  The reason is as follows. Since \(\frac{\partial R^{\prime\prime}(\alpha, \bar p)}{\partial\alpha}=\frac{-1}{\ln2}\frac{(h\sigma_2^2-g\sigma_1^2)\bar p}{(\alpha h\bar p+\sigma_1^2)(\alpha g\bar p+\sigma_2^2)}\leq 0\),  \(R^{\prime\prime}(\alpha, \bar p)\) is monotonically non-increasing with respect to \(\alpha\) over the interval \([0,1]\), and thus attains its maximum at \(\alpha=0\). Now, with the average harvested power constraint added, since the harvested power given in \eqref{eq:harvested energy} in each fading state \(\nu\) is independent of \(\alpha(\nu)\), it is also true that setting \(\alpha^\ast(\nu)=0,\forall\nu\), has no loss of optimality. Combining the above two results, we conclude that \(\alpha^\ast(\nu)=0,\forall\nu\), should be optimal for both problems. Proposition \ref{prop:noCancel-optimal power splitting ratio} is thus proved.

\bibliographystyle{IEEEtran}
\bibliography{TVT_SWIPT}

% Generated by IEEEtran.bst, version: 1.13 (2008/09/30)
\begin{thebibliography}{10}
\providecommand{\url}[1]{#1}
\csname url@samestyle\endcsname
\providecommand{\newblock}{\relax}
\providecommand{\bibinfo}[2]{#2}
\providecommand{\BIBentrySTDinterwordspacing}{\spaceskip=0pt\relax}
\providecommand{\BIBentryALTinterwordstretchfactor}{4}
\providecommand{\BIBentryALTinterwordspacing}{\spaceskip=\fontdimen2\font plus
\BIBentryALTinterwordstretchfactor\fontdimen3\font minus
  \fontdimen4\font\relax}
\providecommand{\BIBforeignlanguage}[2]{{%
\expandafter\ifx\csname l@#1\endcsname\relax
\typeout{** WARNING: IEEEtran.bst: No hyphenation pattern has been}%
\typeout{** loaded for the language `#1'. Using the pattern for}%
\typeout{** the default language instead.}%
\else
\language=\csname l@#1\endcsname
\fi
#2}}
\providecommand{\BIBdecl}{\relax}
\BIBdecl

\bibitem{Zhang2013MIMO}
R.~Zhang and C.~K. Ho, ``{MIMO} broadcasting for simultaneous wireless
  information and power transfer,'' \emph{IEEE Trans. Wireless Commun.},
  vol.~12, no.~5, pp. 1989--2001, May 2013.

\bibitem{Zhou2013SWIPT}
X.~Zhou, R.~Zhang, and C.~Ho, ``Wireless information and power transfer:
  architecture design and rate-energy tradeoff,'' \emph{{IEEE} Trans.Commun.},
  vol.~61, no.~11, pp. 4757--4767, Nov. 2013.

\bibitem{Liu2013}
L.~Liu, R.~Zhang, and K.~Chua, ``{Wireless information transfer with
  opportunistic energy harvesting},'' \emph{{IEEE} Trans. Wireless Commun.},
  vol.~12, no.~1, pp. 288--300, Jan. 2013.

\bibitem{Xu2013multiuser}
J.~Xu, L.~Liu, and R.~Zhang, ``Multiuser {MISO} beamforming for simultaneous
  wireless information and power transfer,'' \emph{{IEEE} Trans. Signal
  Process.}, vol.~62, no.~18, pp. 4798--4810, Sept. 2014.

\bibitem{Wyner1975}
A.~D. Wyner, ``The wire-tap channel,'' \emph{Bell Syst. Tech. J}, vol.~54,
  no.~8, pp. 1355 --1387, Oct. 1975.

\bibitem{Cheong1978}
S.~Leung-Yan-Cheong and M.~Hellman, ``The {G}aussian wire-tap channel,''
  \emph{IEEE Trans. Inf. Theory}, vol.~24, no.~4, pp. 451--456, July 1978.

\bibitem{Goel2008}
S.~Goel and R.~Negi, ``Guaranteeing secrecy using artificial noise,''
  \emph{{IEEE} Trans. Wireless Commun.}, vol.~7, no.~6, pp. 2180--2189, June
  2008.

\bibitem{X.Zhou2010}
X.~Zhou and M.~McKay, ``Secure transmission with artificial noise over fading
  channels: achievable rate and optimal power allocation,'' \emph{{IEEE} Trans.
  Veh. Technol.}, vol.~59, no.~8, pp. 3831--3842, Oct. 2010.

\bibitem{Liao2011transmit_beamforming}
W.~Liao, T.~Chang, W.~Ma, and C.~Chi, ``Qos-based transmit beamforming in the
  presence of eavesdroppers: an artificial-noise-aided approach,'' \emph{IEEE
  Trans. Signal Process.}, vol.~59, no.~3, pp. 1202--1216, Mar. 2011.

\bibitem{Li2013Spatially}
Q.~Li and W.-K. Ma, ``Spatially selective artificial-noise aided transmit
  optimization for {MISO} multi-eves secrecy rate maximization,'' \emph{IEEE
  Trans. Signal Process.}, vol.~61, no.~10, pp. 2704--2717, May 2013.

\bibitem{Liu2014Secrecy}
L.~Liu, R.~Zhang, and K.-C. Chua, ``Secrecy wireless information and power
  transfer with {MISO} beamforming,'' \emph{{IEEE} Trans. Signal Process.},
  vol.~62, no.~7, pp. 1850--1863, April 2014.

\bibitem{bloch2008wireless}
M.~Bloch, J.~Barros, M.~R.~D. Rodrigues, and S.~McLaughlin, ``Wireless
  information-theoretic security,'' \emph{IEEE Trans. Inf. Theory}, vol.~54,
  no.~6, pp. 2515--2534, June 2008.

\bibitem{Barros2006}
J.~Barros and M.~Rodrigues, ``Secrecy capacity of wireless channels,'' in
  \emph{Proc. {IEEE} International Symposium on Information Theory ({ISIT})},
  Seattle, Washington, USA, July 2006, pp. 356--360.

\bibitem{Liang2008}
Y.~Liang, H.~Poor, and S.~Shamai, ``Secure communication over fading
  channels,'' \emph{{IEEE} Trans. Inf. Theory}, vol.~54, no.~6, pp. 2470--2492,
  June 2008.

\bibitem{Khalil2009Opportunistic}
K.~Khalil, O.~O. Koyluoglu, H.~El-Gamal, and M.~Youssef, ``Opportunistic
  secrecy with a strict delay constraint,'' \emph{IEEE Trans. Commun.},
  vol.~61, no.~11, pp. 700--4709, Nov. 2013.

\bibitem{Gungor2013Secrecy}
O.~Gungor, J.~Tan, C.~Koksal, H.~El-Gamal, and N.~Shroff, ``Secrecy outage
  capacity of fading channels,'' \emph{IEEE Trans. Inf. Theory}, vol.~59,
  no.~9, pp. 5379--5397, Sept. 2013.

\bibitem{Gopala2008}
P.~K. Gopala, L.~Lai, and H.~El-Gamal, ``On the secrecy capacity of fading
  channels,'' \emph{{IEEE} Trans. Inf. Theory}, vol.~54, no.~10, pp.
  4687--4698, Oct. 2008.

\bibitem{Khisti2008fading}
A.~Khisti, A.~Tchamkerten, and G.~W. Wornell, ``Secure broadcasting over fading
  channels,'' \emph{Trans. Inf. Theory}, vol.~54, no.~6, pp. 2453--2469, June
  2008.

\bibitem{Koorapaty1998}
H.~Koorapaty, A.~Hassan, and S.~Chennakeshu, ``{Secure information transmission
  for mobile radio},'' in \emph{Proc. {IEEE} International Symposium on
  Information ({ISIT})}, Cambridge, MA,USA, Aug. 1998, p. 381.

\bibitem{Koorapaty2000}
------, ``{Secure information transmission for mobile radio},'' \emph{IEEE
  Communications Letters}, vol.~4, no.~2, pp. 52--55, Feb. 2000.

\bibitem{Yu2006}
W.~Yu and R.~Lui, ``Dual methods for nonconvex spectrum optimization of
  multicarrier systems,'' \emph{{IEEE} Trans. Commun.}, vol.~54, no.~7, pp.
  1310--1322, July 2006.

\bibitem{rockafellar1997convex}
R.~T. Rockafellar, \emph{Convex analysis}.\hskip 1em plus 0.5em minus
  0.4em\relax Princeton university press, 1997.

\bibitem{boyd2004convex}
S.~Boyd and L.~Vandenberghe, \emph{Convex optimization}.\hskip 1em plus 0.5em
  minus 0.4em\relax Cambridge university Press, 2004.

\bibitem{Liu2010MIMOGaussian_broadcast}
R.~Liu, T.~Liu, H.~Poor, and S.~Shamai, ``Multiple-input multiple-output
  {G}aussian broadcast channels with confidential messages,'' \emph{IEEE Trans.
  Inf. Theory}, vol.~56, no.~9, pp. 4215--4227, Sept. 2010.

\end{thebibliography}

\end{document}